\documentclass[11pt]{article}
\usepackage{graphicx}
\renewcommand{\title}[1]{{\noindent\large\bfseries#1\medskip\\}}
\renewcommand{\author}[2]{{\noindent #1 \medskip\\ \small #2 \medskip\\}}
\usepackage[letterpaper,margin=20mm]{geometry}
\usepackage{caption}
\usepackage{subcaption}
\usepackage{comment}
\usepackage{hyperref} 
\usepackage{amsmath,amssymb}
\usepackage{comment}
\usepackage{xcolor}
\usepackage{float}
\usepackage{subcaption}
\usepackage{caption}
\usepackage[export]{adjustbox}
\usepackage{listings}
\usepackage{cancel}
\usepackage{multirow}
\newcommand{\E}{\mathbb{E}}
\begin{document}

\title{Beyond the Wisdom of the Crowd: How Network Topology Distorts Collective Perception}

\author{
Giovanni Palermo,\textsuperscript{1, 2}
Vittorio Loreto,\textsuperscript{1,3,2,5}
Giulio Cimini,\textsuperscript{4,2}
}
{
1. Sapienza University of Rome, Physics Department, Piazzale A. Moro 2, 00185 Rome (Italy)\\
2. Enrico Fermi Research Center, Via Panisperna 89/A, 00184 Rome (Italy)\\
3. Sony Computer Science Laboratories - Rome, Joint initiative CREF-SONY, Enrico Fermi Research Center, Via Panisperna 89/A, 00184 Rome (Italy)\\
4. University Tor Vergata, Physics Department and INFN, 00133 Rome (Italy)\\
5. Complexity Science Hub, Metternichgasse 8, 1030, Vienna (Austria)
}

\bigskip

\begin{abstract}

Cognitive biases are often attributed to heuristics or limited information. Yet the structure of social networks is a key, often-overlooked source of perceptual bias. When information passes through social connections, the network alone can systematically distort how individuals view society. We use a simple model in which agents have a binary attribute (e.g., atheist or believer) and show that network topology alone can cause misperceptions of peers' attributes. These misperceptions persist even after aggregation and challenge the idea of the "wisdom of the crowd." We derive an estimator that predicts the size and direction of these biases from network features. We validate our findings using three large-scale opinion surveys. Our results show that network structure is a critical factor in collective perception, with major implications for reducing segregation, polarisation, and the marginalisation of minorities.

\end{abstract}

\section{Introduction}

Over the past decades, the social and behavioural sciences have shifted profoundly. Researchers have set aside the classical view of humans as fully rational agents, capable of processing all information and computing optimal choices, in favour of a more realistic account of decision-making under cognitive and informational constraints. Simon’s seminal work~\cite{Simon1955} introduced \textit{bounded rationality}—the concept that humans make decisions using limited cognitive resources and imperfect information. He suggested that people do not evaluate all possible options but instead stop searching once they find one that is “good enough” (a process known as satisficing).

Subsequent research by Kahneman and Tversky~\cite{Tversky1974} demonstrated that humans use cognitive shortcuts, or heuristics—mental strategies or rules of thumb—that systematically deviate from rationality. People display biases such as overconfidence~\cite{Zhang2024}, anchoring~\cite{Tversky1974, Chapman1999}, and confirmation bias~\cite{Nickerson1998, Klayman1987} as consistent patterns rather than random errors. Human cognition evolved to maximise efficiency rather than accuracy~\cite{Todd1999}.

However, cognitive biases alone cannot fully account for the collective patterns we observe at larger scales. Beliefs, decisions, and errors spread through social ties, as individuals both influence and are influenced by others. Network science~\cite{Cohen2002, BarabasiPosfai2016,Newman2018} offers powerful tools to analyse how such heuristics and biases diffuse across groups, shaping opinion formation, collective behaviour, and the spread of (mis)information. Importantly, social connectivity can both result from and reinforce cognitive biases.

A prominent example is the emergence of {\em echo chambers}—social environments where individuals are primarily exposed to opinions similar to their own—in online platforms~\cite{DelVicario2016, Cinelli2021}. Users in these spaces connect preferentially with like-minded others, often driven by cognitive tendencies like confirmation bias—which favours information supporting existing beliefs~\cite{Gallo2020}—or reinforcement dynamics, whereby repeated exposure to similar views increases their perceived validity~\cite{Brugnoli2021}. Such mechanisms amplify misperceptions by filtering out dissenting perspectives and exposing individuals only to supportive content. For instance, people often overestimate the popularity of their political views~\cite{Ross1977, Mullen1985, Krueger1994, Krueger2007}, largely because their contacts share similar opinions—an effect observed across online networks~\cite{Conover2021, viceJournalistsTrump}. Additionally, Lee et al.~\cite{Lee2019} described how homophily (the tendency to associate with similar others) and minority-group size can together produce perception biases: both over- and underestimation of minority-group size can emerge solely from structural properties of social networks.

In this work, we focus on these {\em network-induced perception biases}, defined as systematic distortions in users' perceptions which emerge from the structure (i.e., the pattern of connections) in social networks, even when individuals are themselves free of cognitive bias. We argue that users extract more information from the network than merely their neighbours’ direct opinions. We extend Lee's model to account for global network effects, clarifying which specific network attributes contribute to perception biases.

We examine how these structural biases distort the {\em{wisdom of the crowd}}~\cite{GALTON1907}, the idea that combining many independent judgments—each person offering their best answer uninfluenced by others—yields a more accurate estimate than relying on any single opinion. This principle grounds recent polling innovations, such as the "election-winner"~\cite{Graefe2014, Murr2019, LewisBeck1989} and "social-circle"~\cite{Galesic2018,BruinedeBruin2022} questions, which gather perceptions from respondents and their contacts. While these methods show promise, correlated responses (when answers are not independent) or biased judgments (when they systematically miss the true value) can compromise their accuracy~\cite{DavisStober2014, Lorenz2011, Becker2017, Kameda2022}.

Here, we show that network topology alone can generate perception biases and undermine the wisdom of the crowd. We use a simple model to generate graphs with two communities (e.g., believers and atheists), allowing precise control over community size and interconnectivity. 
We derive user perceptions by running a message-passing algorithm over network links, revealing how distorted perceptions emerge endogenously. 
Crucially, these biases persist even when perceptions are averaged across the entire network, undermining the wisdom of the crowd. 
We identify key structural determinants of the distortion, including community degree, size imbalance, and polarisation. Finally, we derive an analytical expression that predicts the population-level bias and validate it using survey data from~\cite{Lee2019}.

Overall, our findings demonstrate that collective perception is shaped jointly by cognition and network architecture. Recognising these topological sources of bias can help explain persistent societal misperceptions—such as political belief gaps between groups~\cite{perceptiongapPerception}—and improve collective forecasting in settings where aggregation fails due to network-driven distortions.

\section{Results}

\subsection{Theoretical and simulation framework}

We model individuals’ social connections 
using a graph with $N$ nodes, where each node represents an individual and links indicate social ties. Each node $i$ is assigned a fixed attribute $s_i$, treated as a binary variable taking values $s_i = \pm 1$. This attribute may correspond to a binary characteristic, such as habit (smoker or non-smoker), political orientation (Democrat or Republican), religious affiliation (believer or atheist), or sex (male or female). The population includes $N_+$ nodes with $s = +1$ and $N_- = N - N_+$ nodes with $s = -1$. Nodes with the same $s$ form a "community". The \emph{mean attribute} across the population is: 
\begin{equation}
    m=\frac{1}{N}\sum_{i=1}^N s_i=\frac{N_+-N_-}{N}
    \label{eq.magnetization}
\end{equation}
taking values in $[-1, 1]$ and quantifying the balance between the two groups: $m = 0$ indicates equal-sized communities, while $m > 0$ ($m < 0$) implies that the $s = +1$ ($s = -1$) attribute is predominant. Given this setup, our goal is to quantify how individuals perceive the mean attribute in the population. For instance, if the attribute distinguishes smokers from non-smokers, we aim to simulate how a person would answer the question: “What fraction of people in your country are smokers?” 

\paragraph*{Graph generation.} 
We generate the social graph using a Stochastic Block Model (SBM)~\cite{Holland1983} with two communities, corresponding to the groups of individuals holding attribute $+1$ and $-1$, respectively. In this model, the probability $p$ determines the likelihood of links forming between individuals from different groups (inter-group connections), while $k_+$ and $k_-$ specify the average number of connections within the $+1$ and $-1$ groups (intra-group connections), respectively. See Methods for further details. 

The rationale for adopting the SBM framework is tied to our intuition about social networks. We expect that homophily—the tendency of individuals to connect with others who are similar to them~\cite{McPherson2001}—plays a key role in shaping personal perceptions. A node embedded in a community of like-minded individuals (i.e., sharing the same attribute) is more likely to develop a biased view of the prevalence of alternative opinions in the network. 

\begin{figure}[p]
    \centering
\includegraphics[width=\linewidth]{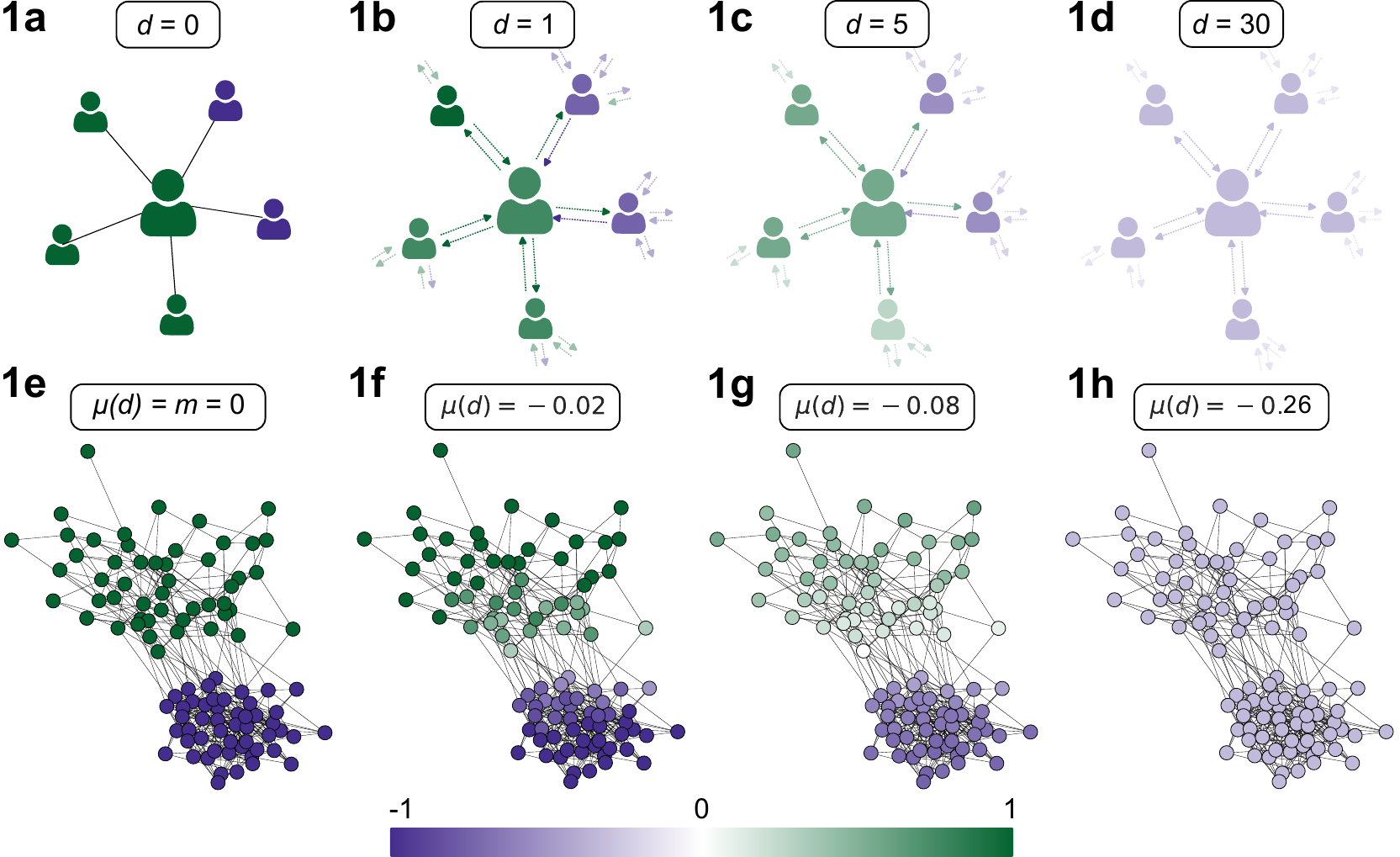}
    \caption{\textbf{Visual representation of the message-passing algorithm.} (1a) The simple graphs in the first row show how perception forms for one user at the centre and their social connections. At $d=0$, nodes do not know about others, so each only knows its own attribute: $\mu_i(0)=s_i$ ($+1$: green, $-1$: purple). (1b) At $d=1$, each user shares their own attribute with their nearest neighbours, resulting in a revised perception of the attribute balance in the network (indicated by the composite user colour between green and purple). As this process is iterated over greater distances, each user continues to receive messages from neighbours, enabling the focal user to obtain information from nodes farther within the network. (1c) After $d=5$ iterations, the user has received messages from peers up to five steps away. (1d) At $d=30$, the user attains a stable perception of the average attribute in the network. (1e) The second row shows the process for the whole network, using a stochastic block model (SBM) with $N=100$, $m=0$, $k_+=5$, $k_-=10$, and $p=0.01$. At $d=0$, no propagation has occurred, so each node's perception corresponds to its own attribute. (1f) At $d=1$, each node only gets messages from its neighbours, and nodes with neighbours in both communities develop a milder perception of the mean attribute. (1g) At $d=5$, signals have reached nearly all nodes in the network. (1h) At $d=30$, perceptions get close to stable values, which are not the same as m (the final average is $-0.26$, not $0$), challenging the idea of the wisdom of the crowd. This happens because $-1$ nodes have more connections, so they share their views more widely.}
    \label{fig1}
\end{figure}

\paragraph*{Message passing.} We compute the perception of each node about the mean attribute of the network using a message-passing algorithm (Figure~\ref{fig1}). Our idea is that information about others' attributes flows through the graph's links—each individual receives signals from their neighbours, who in turn are influenced by their own contacts, and so on.  To simulate this process, we model perception as beliefs that form according to the DeGroot model~\cite{Degroot1974}: the perception of node $i$, $\mu_i$, is given by the average of her neighbours' perceptions. Such a fixed-point solution can be reached iteratively using network propagation steps $d$.  Initially, at $d=0$, nodes have no information at all on others, so their perception is their own attribute: $\mu_i(0)=s_i$. At $d=1$ nodes can get information from their neighbours, so their perception $\mu_i(1)$ is simply given by the average attribute of the nearest neighbours—i.e., the \emph{social circle}. In the next step, $d=2$, nodes' perceptions still form from those of their neighbours, which are now determined by their own social circles; in this way, nodes also receive attribute information from their neighbours' neighbours (nodes at distance $d=2$). By iterating, the information travels longer distances $d$. Overall, one has:
\begin{equation}
    \mu_i(d) = \frac{\sum_{j\in{\cal N}_i} \mu_j(d-1)}{k_i}
    \label{eq.perception}
\end{equation}
where $d$ is the signal traveling distance, ${\cal N}_i$ is the set of first neighbors of $i$ and $k_i$ is the number of such neighbors (the degree of $i$). As $d$ grows, perception reaches its stationary value (the fixed-point solution of eq.~\eqref{eq.perception}).

\begin{figure}[p]
    \centering
    \includegraphics[width=\linewidth]{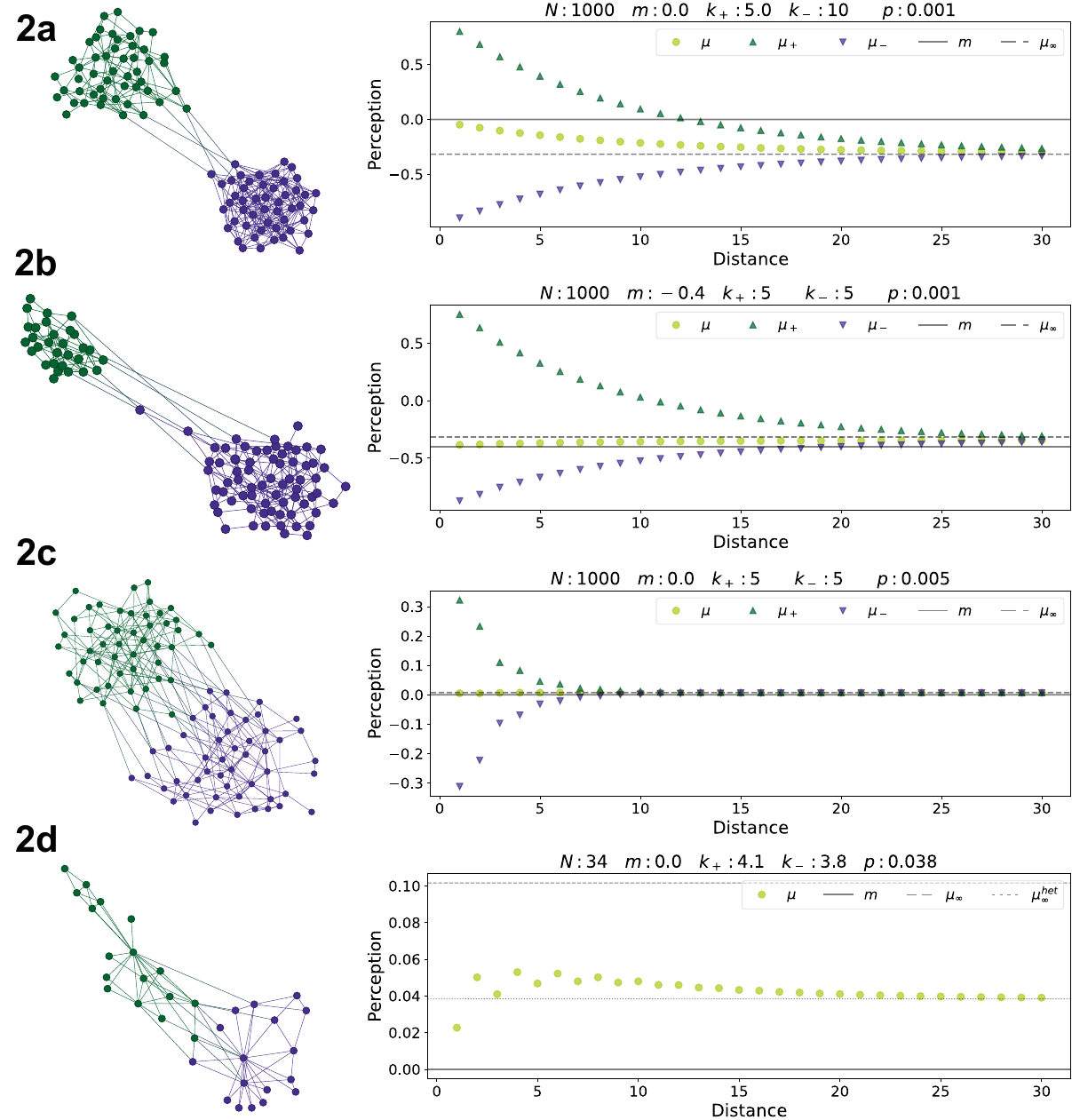}
    \caption{\textbf{Message passing simulations in four different scenarios.} Plots on the right column show the mean attribute $m$ (continuous line) and the stationary perception $\mu_\infty$ (dashed line), while symbols represent the mean perception of the whole network ${\mu}$ and the mean perception within each community ${\mu}_{\pm}$ as a function of distance $d$. The perception bias is given by the difference between $m$ and $\mu(d)$ for $d\to\infty$. All graphs on the left are merely illustrative of the real ones on which the simulations were run. (2a) SBM with same community size ($N_+=N_-$), small mixing (low $p$) and different internal degrees $k_+\neq k_-$; (2b) SBM with different community size ($N_+\neq N_-$), small mixing (low $p$) and same internal degrees $k_+= k_-$; (2c) SBM with equal community size ($N_+=N_-$), high mixing (high $p$) and same internal degrees $k_+= k_-$. (2d) Karate club graph with affiliation attributes. Here, the dotted line is the stationary perception computed with heterogeneous mean-field approximation, i.e., accounting for degree correlations in the real network. }
    \label{fig2}
\end{figure}

\paragraph*{Perception bias.} 
We focus on the \emph{mean perception} of the network, given by $\mu(d)=\sum_{i=1}^N \mu_i(d)/N$. We are interested in understanding whether the population-average perception matches the mean attribute $m$ of eq.~\eqref{eq.magnetization}. It turns out that in general this is not the case, as evident from the stationary value of the mean perception—derived analytically using mean-field approximation at the level of communities (see Methods and Supplementary Information S1):
\begin{equation}
    \mu_{\infty} = \E \Big[\lim_{d\to\infty}\mu(d)\Big]=\frac{k_+N_+ - k_-N_-}{k_+N_+ + k_-N_- +4N_+N_-p}
    \label{eq:m_th}
\end{equation}
This formula shows that the network's features intertwine to generate the mean perception. The difference between $\mu$ and $m$ quantifies the {\em perception bias} due to purely topological effects, which we discuss below and illustrate in Figure~\ref{fig2}.

\paragraph*{Connectivity heterogeneity.}
When communities have equal size ($N_+=N_-$), as in Fig.~\ref{fig2}a, then by definition $m=0$ and

\begin{equation}
    \mu_{\infty} = \frac{k_+ - k_-}{k_+ + k_- +2Np}
    \label{eq:m_th_m_zero}
\end{equation}

This value matches the mean attribute only if $k_+=k_-$; if instead $k_+ > k_-$, the perception shifts towards $+1$ (and viceversa). Hence, the higher the internal degree of a community, the higher its impact on the aggregate perception, as its nodes have more channels to dispatch messages to others. As a result, communities with higher (lower) connectivity are perceived as bigger (smaller) than they actually are. This effect is reminiscent of the {\em friendship paradox}~\cite{Feld1991} taking place at the community level.

\paragraph*{Community-size imbalance.}
When communities have different sizes, as in Fig.~\ref{fig2}b, we have $m\neq 0$. Consider the simple situation $k_+=k_-=k$, in which:

\begin{equation}
    \mu_{\infty} = 
    m\left(1+\frac{1-m^2}{2k}Np\right)^{-1}
    \label{eq:m_th_k_zero}
\end{equation}

We see that the absolute value of $\mu_{\infty}$ is always smaller than the absolute value of $m$ (unless $m=0$), since $m^2<1$. Thus, the majority (minority) is perceived on average as smaller (bigger) than it actually is. This happens because the average degree of nodes in the two communities is $k+2N_-p$ and $k+2N_+p$, respectively, hence the larger (smaller) community has a lower (higher) degree and sends fewer (more) messages.

\paragraph*{Polarisation.} 
The probability of connection $p$ between nodes in different communities serves as a homophily or polarisation parameter: when $p$ is small (large), communities have a low (high) level of mixing. 
Polarisation has opposite effects in the two previous cases of connectivity heterogeneity and community-size imbalance. In the first case, when $m=0$, a large $Np$ term in eq.~\eqref{eq:m_th_m_zero} makes the denominator larger, mitigating the effect of degree heterogeneity. Thus, high mixing between communities reduces perception bias (and vice versa). If, instead, the two communities are imbalanced and $k_+=k_-$, a small $Np$ term in eq.~\eqref{eq:m_th_k_zero} allows perception to approach $m$. Thus, low mixing helps to mitigate perception bias. Polarisation also has another implication: high values delay reaching the stationary state to longer distances, while low values speed up the process (Fig.~\ref{fig2}c). Additionally, for very high polarisation, the message-passing might not reach the stationary state at all, and each community might hold a biased perception of the other, perceiving it as smaller than it actually is. This effect can be seen by looking at the \emph{community mean perception} $\mu_\pm(d)=\sum_{i: s_i=\pm1} \mu_i(d)/N_\pm$ in Figure~\ref{fig2}; before reaching the stationary state, each community overestimates its own size, as its perception is systematically biased towards its own attribute.

\paragraph*{Karate club network.} We further test the model on the well-known Karate club network~\cite{Zachary1977}, where the $N=34$ nodes represent club members and links represent interactions between members outside the gym. After a conflict between the administrator and instructor, the graph splits into two communities. This division leads to perception bias in our simulations (Fig.~\ref{fig2}d). In this case, however, eq.~\eqref{eq:m_th} does not accurately capture stationary perception. This is because $k_i$ and $s_i$ values of individual nodes are correlated, in contrast to the SBM case. We therefore use a heterogeneous mean-field approach (see Methods), grouping nodes into degree classes and thus faithfully predicting the stationary perception. Note, however, that knowing the correlation between degrees and attributes is required.

\paragraph*{Effects on the wisdom of the crowd}
In all cases considered, the mean perception of individuals does not match the true mean of the network's attribute. The principle of the \emph{wisdom of the crowd} states that aggregating many individual estimates should yield an accurate collective prediction—if individual judgments are unbiased. However, the network-induced distortions discussed above introduce systematic biases that undermine this effect. For example, imagine an external observer estimating the fraction of smokers in a population by aggregating individuals’ perceptions. Due to the network effects described, the aggregated estimate would be incorrect because $\mu$ generally differs from $m$. This phenomenon is observed in reality: Lee et al.~\cite{Lee2019} found that people’s perceptions of social issues consistently deviate from reality and linked this shift to a social-circle effect. In the next subsection, we show that the message-passing algorithm more accurately accounts for collective misperception than the social circle.

\subsection{Real data}
We test our predictions on survey data by Lee et al., who interviewed $100$ people in Germany, the USA, and South Korea on several issues ($10$ in Germany and the USA, $7$ in South Korea). For each issue, they asked participants about their own binary attribute, their estimate of the fraction of their social acquaintances with that attribute, and their estimate of the fraction of the country with that attribute. For example, they asked each person: “Are you a smoker?”, “What is the percentage of smokers in your social circle?”, “What is the percentage of smokers in your country?''. In our framework, these answers map to each user's own attribute, her social circle, and her perception of the mean attribute, respectively.

Lee et al. showed that the misperception of group size—how many smokers people think there are in their country—correlates with their social circle (i.e., how many smokers they have in their social circle). This also applies to our framework: a high level of homophily (preference for associating with similar individuals) leads to networks with strong community splitting. In such networks, messages from one community must travel farther to reach another. Here, we aim to better estimate perception bias by focusing on what happens beyond the social-circle level (where distance $d>1$ between individuals). To this end, we rewrite the average perception of eq.~\eqref{eq:m_th} using the social circle values of the communities (see Supplementary Information S2 for the full derivation):

\begin{equation}
    \mu_\infty = \frac{\mu_+(1) + \mu_-(1)}{2-\mu_+(1) + \mu_-(1)}
    \label{eq:mu_sample}
\end{equation}

This expression can be readily evaluated using the survey data at our disposal, without requiring knowledge of the network's topological features nor the mean attribute $m$.

\begin{figure}[H]
    \centering    \includegraphics[width=\linewidth]{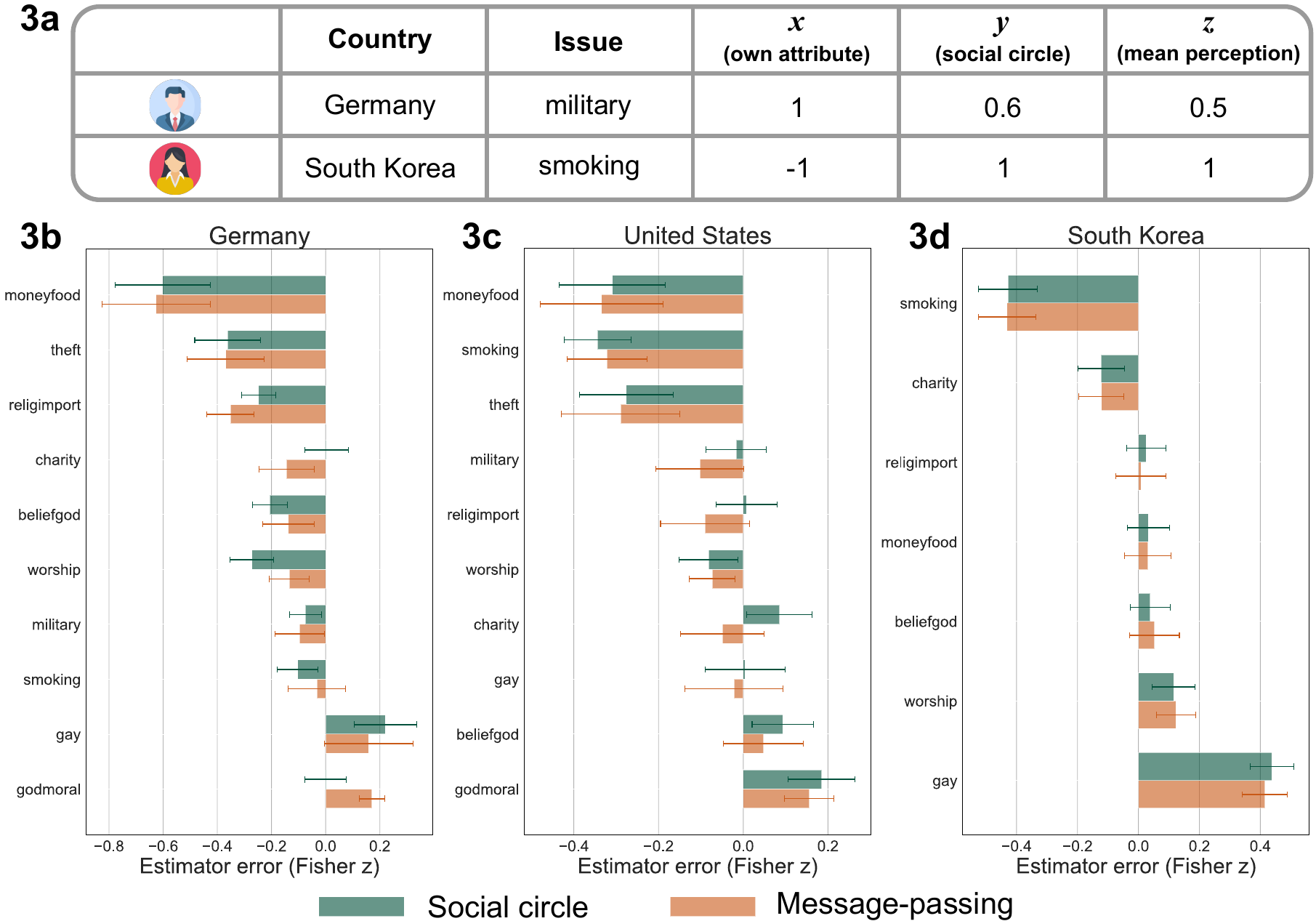}
    \caption{
    \textbf{Assessment of the perception bias estimators.} (3a) Sample illustration of survey data. For each user, country, and issue, the answers include her attribute $x_i$, her social circle $y_i$, and her country-level perception $z_i$. (3b-d) Error of the social circle estimator $\hat{\mu}_1$ and the message-passing estimator $\hat{\mu}_\infty$  in predicting the average perception $\hat{\mu}$ in the surveys, for the three countries in the data. The error bars account for one standard deviation of the estimator. Values are reported after the Fisher z-transform, and issues are ordered by message-passing error. }
    \label{fig3}
\end{figure}

\paragraph*{Perception bias estimators.}
We denote survey information for each user $i$ as her declared attribute $x_i$, her social circle $y_i$, and her country-level perception $z_i$ (see Figure~\ref{fig3}a for an example). Assuming that participants reported these values accurately and that the answers are uncorrelated, the average perception of the population in each country can be inferred from $z$. For each issue and country, we split the sample into two subsamples based on the value of the binary variable $ x = \pm 1$. We then compute the sample means of $z$ in the two sub-samples and denote them as $\overline{z}_\pm$. We know the presumed real value of $m$, which is provided with the data and comes from previous studies. Therefore, we post-stratify the estimation of $\mu$, calculating it as the weighted mean of $\overline{z}_+$ and $\overline{z}_-$, where the weights are the relative size of the two communities, namely $n_\pm=\frac{1\pm m}{2}$, obtaining our estimation of the average perception:

\begin{equation}
    \hat\mu = n_+\overline{z}_+ + n_-\overline{z}_-
\end{equation}

Similarly, we compute the sub-sample means of $y$, leading to our estimation of the social circle values $\hat{\mu}_\pm(1)=\overline{y}_\pm$, and get the social circle estimator $\hat{\mu}_1$:

\begin{equation}
    \hat{\mu}_1 = n_+\overline{y}_+ + n_-\overline{y}_-
    \label{eq:mu_1}
\end{equation}

Instead the message-passing estimator $\hat{\mu}_\infty$ is obtained by plugging the $\hat{\mu}_\pm(1)=\overline{y}_\pm$ values into eq. \eqref{eq:mu_sample}:

\begin{equation}
    \hat{\mu}_\infty= \frac{\overline{y}_+ + \overline{y}_-}{2- \overline{y}_+ + \overline{y}_-}
\end{equation}

%


%

\paragraph*{Message-passing vs Social circle.}
We now compare the predictive power of our message-passing estimator $\hat{\mu}_\infty$ and of the social circle $\hat{\mu}_1$. Since the actual perception $\hat{\mu}$ and the predicted values are bounded between $-1$ and $1$, we apply the Fisher z-transformation~\cite{Fisher1915} before computing errors and uncertainties, so that the analysis is performed in an approximately unbounded and Gaussian-like space where error propagation and statistical testing are more reliable (see Supplementary Information S3). Figures~\ref{fig3}b-d report the prediction errors of the two estimators, namely $\hat{\mu}_\infty-\hat{\mu}$ and $\hat{\mu}_1-\hat{\mu}$, for each country and issue, along with their associated uncertainty (we remand to Supplementary Information S4 for the computation of such uncertainties).

As expected, the errors of the estimators generally correlate, since the message-passing value is based on the social-circle value. To assess which estimator better captures the average perception measured by the surveys, we evaluate the reduced $\chi^2$ statistic, which measures how well the predictions agree with the observations relative to their uncertainties. An estimator with reduced $\chi^2 \sim 1$ is consistent with the data within its uncertainty bars, while values much larger than $1$ indicate systematic deviations or underestimated uncertainties (see Supplementary Information S5 for further information).

Table~\ref{tab:results} shows that the reduced $\chi^2$ is smaller for the message-passing estimator than the social circle one for all the countries in the survey. Also, for all countries, the reduced $\chi^2$ is larger than $1$. This is not due to underestimation of error bars, which, as shown in Figure~\ref{fig3}b-d, are quite large due to the small sample size. Indeed, South Korea, with just $7$ issues, has a significantly higher $\chi^2$ statistic. A plausible explanation for the discrepancy between perception and prediction is that survey responses arise from the interplay between cognitive biases and network effects, whereas the estimators discussed here are sensitive only to the latter. Nevertheless, we can conclude that, for this survey, perception bias is better predicted by the message-passing estimator; newer, larger studies may confirm these findings.

\setlength{\tabcolsep}{8pt} 
\renewcommand{\arraystretch}{1.5} 

\begin{table}[!http]
\centering
\begin{tabular}{clccc}
\hline
\multicolumn{1}{|c|}{} &
  \multicolumn{1}{c|}{\textbf{Overall}} &
  \multicolumn{1}{c|}{\textbf{Germany}} &
  \multicolumn{1}{c|}{\textbf{United States}} &
  \multicolumn{1}{c|}{\textbf{South Korea}} \\ \hline
\multicolumn{1}{|c|}{} &
  \multicolumn{1}{c|}{\textbf{$\chi^2_{27}$}} &
  \multicolumn{1}{c|}{\textbf{$\chi^2_{10}$}} &
  \multicolumn{1}{c|}{\textbf{$\chi^2_{10}$}} &
  \multicolumn{1}{c|}{\textbf{$\chi^2_{7}$}} \\ \hline
\multicolumn{1}{|c|}{\textbf{Social circle $\hat\mu_1$}} &
  \multicolumn{1}{c|}{4.8} &
  \multicolumn{1}{c|}{4.6} &
  \multicolumn{1}{c|}{3.4} &
  \multicolumn{1}{c|}{7.1} \\ \cline{1-5}
\multicolumn{1}{|c|}{\textbf{Message-passing $\hat\mu_\infty$}} &
  \multicolumn{1}{c|}{4.0} &
  \multicolumn{1}{c|}{3.6} &
  \multicolumn{1}{c|}{2.6} &
  \multicolumn{1}{c|}{6.7} \\ \hline
\end{tabular}
\caption{\textbf{Statistical analysis of the estimator errors}. Values of the reduced $\chi^2$ for the social circle and the message-passing estimator. The subscript under the $\chi^2$ indicated the degrees of freedom of the test, which in this case corresponds to the number of issues for each country.}
\label{tab:results}
\end{table}

\section{Discussion}
Cognitive biases are typically described as systematic errors arising from humans’ imperfect use of information. These have been described as adaptive heuristics or evolutionary shortcuts that enable quick decisions under uncertainty. In this work, we have identified another contribution to the misperception of group sizes, namely the structure of the network of social connections, showing how group-based perception biases can arise from social network features.

We investigate a setting where users possess a binary attribute and perceive others' attributes through social acquaintances. Social ties decisively shape information flows and serve as selective filters, systematically distorting perceptions. We demonstrate that these network-induced biases emerge independently of cognitive or informational errors and arise solely from network topology.

These findings have important implications. Real social networks exhibit group-size imbalance, heterogeneous connectivity, and polarisation~\cite{Conover2021, viceJournalistsTrump}. We demonstrated that these features alone can generate systematic misperceptions of group sizes. This mechanism helps explain empirical observations such as the overestimation of immigrant populations~\cite{ipsosIpsos} or minority groups~\cite{leurispesAntisemitismoPregiudizi}.
This can happen when a community is weakly connected and thus perceived as smaller than others. On the other hand, a small but strongly connected minority may be perceived as bigger by others. Additionally, high polarisation reduces interactions between communities, limiting the flow of information between them.
Taken together, these network-driven distortions may contribute to the \emph{false consensus}~\cite{Ross1977} and \emph{false uniqueness}~\cite{suls2007} effects, in which people perceive their own attributes as more diffuse or rare than they actually are.


Such distortions have clear policy relevance. Network-induced misperceptions can reinforce segregation, amplify prejudice, and skew public opinion on social issues. Populations may perceive themselves as poorer, less safe, or less healthy than they are, undermining trust in institutions and obscuring the effects of effective policies. Reducing structural segregation and polarisation by promoting inter-group connectivity could therefore mitigate the formation of false collective beliefs.

Our framework also  provides insights into the “wisdom of the crowd.” Aggregating individual perceptions across a network does not necessarily recover the true population value. Not only are the groups individually biased, but averaging over the whole network also does not yield the true mean as an outcome. This effect has already been shown at the social-circle level~\cite{Palermo2025} and now has strong confirmation in the context of longer-range interactions.

We also verified the robustness of our findings across alternative message-passing algorithms (Supplementary Information S6) and network models, including homophilic Barabási–Albert graphs~\cite{Lee2019} (Supplementary Information S7). Future work should extend this framework to multi-valued or continuous attributes and combine it with controlled experiments designed to disentangle network-induced effects from cognitive biases or exogenous information. Larger and more diverse samples would also help confirm the superior performance of our message-passing estimator over the social-circle approach in predicting perception bias.

In sum, our results show that the structure of social connections can independently generate systematic perception biases and affect the accuracy of collective judgments. Survey data from three countries confirm these mechanisms in real settings, supporting our model’s predictive power. Beyond theoretical insight, this framework offers a practical lens for identifying and mitigating distorted perceptions in the population, enabling policymakers to understand and prevent perception biases on socially relevant issues.

\section{Methods}

\paragraph*{Stochastic Block Model (SBM).}
We generate undirected SBM networks using the Python \texttt{igraph} library by tuning the following parameters: $N$ number of nodes; $m$ mean attribute; $k_\pm$ average internal degree of nodes in the two communities; $p$ probability of connection between nodes in different communities. We then assign attribute $s=+1$ to the $N_+=N(1+m)/2$ nodes of the community with internal connectivity $k_+$, and attribute $s=-1$ to the $N_-=N-N_+$ nodes of the community with internal connectivity $k_-$. The average external degree of the $\pm1$ community is $2pN_\mp$, and the degree distribution is binomial for both the internal and external degrees.

\paragraph*{Community mean-field.} To compute the expected stationary perception $\mu_\infty$ of eq. \eqref{eq.perception} we use a mean-field approximation at the level of communities, meaning that we assume nodes in the same community ($\pm1$) have the same expected value of perception ($\E[\mu_\pm]$). At the propagation step $d=1$, perception is simply the average of the neighbours' attributes. We expect each node in the $\pm 1$ community to have $k_\pm$ neighbours with $s=\pm1$ and $2pN_\mp$ neighbours with $s=\mp1$. We get that the expected perception of nodes in the two communities is:

$$\E[{\mu}_+(1)] = \frac{k_+(+1) + 2pN_-(-1)}{k_+ + 2pN_-} \qquad
    \E[{\mu}_-(1)] = \frac{k_-(-1) + 2pN_+(+1)}{k_- + 2pN_+}$$

given that $\E[\mu_\pm(0)] = \pm1$.\\
For $d>1$, perception is given by the average of the neighbours' perceptions. On average, each node in the $\pm1$ community has $k_\pm$ neighbors with ${\mu}_\pm(d)$ and $2pN_\mp$ neighbors with ${\mu}_\mp(d)$, so we get a system of two iterative equations:

\begin{equation}
    \E[{\mu}_+(d+1)] = \frac{k_+\;\E[{\mu}_+(d)] + 2pN_-\;\E[{\mu}_-(d)]}{k_+ + 2pN_-} \qquad
    \E[{\mu}_-(d+1)] = \frac{k_-\;\E[{\mu}_-(d)] + 2pN_+\;\E[{\mu}_+(d)]}{k_- + 2pN_+}
    \label{eq:recurrence_system}
\end{equation}

The system forms a pair of linear recurrence relations with constant coefficients, admitting a closed-form solution through standard techniques for linear difference equations.\\
For $d\to\infty$, the two values converge to the same stationary solution given by eq. \eqref{eq.perception}.

\paragraph*{Heterogeneous mean-field.}
When node degrees and attributes are correlated, the previous approach does not accurately predict the stationary perception. Therefore, we adopt a heterogeneous mean-field framework, where we assume that all nodes with the same degree $k$ share the same expected perception value, denoted as $\E[{\mu}_k]$. Given the probability distribution $p(k’|k)$ for the degree $k’$ of a neighbor of a node with degree $k$, we can write:

$$\E[{\mu}_k(d+1)] = \sum{k’} \E[{\mu}_{k’}(d)] p(k’|k) = \sum_{k’} \E[{\mu}_{k’}(d)] \frac{k’p(k’)}{\langle k\rangle}$$

where we used the uncorrelated network approximation $p(k’|k)=k’p(k’)/\langle k\rangle$.
Note that since there is no dependence on $k$ in the right-hand side of the above expression, convergence occurs  after one iteration ($d=1$) to the value:

$$\mu_\infty^{het} = \sum_{k’} s_{k’}\frac{k’p(k’)}{\langle k\rangle}=\frac{\sum_i{k_is_i}}{\sum_i{k_i}}$$

which depends on how the binary attribute $s$ is distributed across nodes with different degrees $k$.

\paragraph*{Acknowledgements.}
We sincerely thank Professors Fariba Karimi and Eun Lee for providing the survey data used in this study.

{\small
\bibliographystyle{naturemag}

}

\clearpage

\setcounter{section}{0}
\renewcommand{\thesection}{S\arabic{section}}
\setcounter{figure}{0}
\renewcommand{\thefigure}{S\arabic{figure}}
\setcounter{table}{0}
\renewcommand{\thetable}{S\arabic{table}}

{\LARGE \bfseries Supplementary Information \\[1.5em]}

{\large \bfseries  \noindent Beyond the Wisdom of the Crowd: How Network Topology Distorts Collective Perception \\[1em]}

\author{
Giovanni Palermo,\textsuperscript{1, 2}
Vittorio Loreto,\textsuperscript{1,3,2,5}
Giulio Cimini,\textsuperscript{4,2}
}
{
1. Sapienza University of Rome, Physics Department, Piazzale A. Moro 2, 00185 Rome (Italy)\\
2. Enrico Fermi Research Center, Via Panisperna 89/A, 00184 Rome (Italy)\\
3. Sony Computer Science Laboratories - Rome, Joint initiative CREF-SONY, Enrico Fermi Research Center, Via Panisperna 89/A, 00184 Rome (Italy)\\
4. University Tor Vergata, Physics Department and INFN, 00133 Rome (Italy)\\
5. Complexity Science Hub, Metternichgasse 8, 1030, Vienna (Austria)
}

\bigskip

\section{Derivation of stationary theoretical solution}

We derive the stationary theoretical solution $\mu_\infty$ from the system (10), contained in the Methods section in the main text.
To simplify the notation, we rename $\E[\mu_+]$ as $x$ and $\E[\mu_-]$ as $y$, so the system becomes

\[
\begin{aligned}
x(d+1) &= \frac{k_+\,x(d) + 2pN_-\,y(d)}{k_+ + 2pN_-},\\[6pt]
y(d+1) &= \frac{k_-\,y(d) + 2pN_+\,x(d)}{k_- + 2pN_+},
\end{aligned}
\]

with the initial conditions $x(0)=1,\; y(0)=-1$.

To solve the linear recurrence system, we seek two constants \(\alpha,\beta>0\) such that
\[
I(d)=\alpha\,x(d)+\beta\,y(d)
\]
is constant in \(d\).  We impose
\[
\alpha\,x(d+1)+\beta\,y(d+1)=\alpha\,x(d)+\beta\,y(d).
\]
Substituting the update rules, one finds that the choice
\[
{\;\alpha = N_+\big(k_+ + 2pN_-\big),\qquad
\beta = N_-\big(k_- + 2pN_+\big)\;}
\]
makes the identity hold for all \(d\). Hence \(I(d)\) is a conserved quantity for each distance.

At \(d=0\):
\[
I(0)
= \alpha x(0) + \beta y(0)
= N_+\big(k_+ + 2pN_-\big)\cdot 1
  + N_-\big(k_- + 2pN_+\big)\cdot(-1).
\]
The \(2pN_+N_-\) terms cancel, giving
\[
I(0) = N_+k_+ - N_-k_-.
\]

As \(d\to\infty\), the system reaches consensus:
\[
x(d),y(d)\longrightarrow \mu_\infty.
\]
Thus
\[
I(\infty)
= \alpha\mu_\infty + \beta\mu_\infty
= \mu_\infty(\alpha+\beta).
\]

Conservation gives \(I(0)=I(\infty)\), hence
\[
\mu_\infty(\alpha+\beta)=N_+k_+ - N_-k_-.
\]

Compute
\[
\alpha+\beta
= N_+\big(k_+ + 2pN_-\big)
+ N_-\big(k_- + 2pN_+\big)
= N_+k_+ + N_-k_- + 4pN_+N_-.
\]

Therefore,
\[
{
\mu_\infty
= \frac{N_+k_+ - N_-k_-}
       {\,N_+k_+ + N_-k_- + 4pN_+N_-\,}
}
\]

and consequently,
\[
\lim_{d\to\infty} x(d) = \lim_{d\to\infty} y(d) = \mu_\infty.
\]

\section{Accuracy of theoretical estimation}

\begin{figure}[!h]
    \centering
    \includegraphics[width=\linewidth]{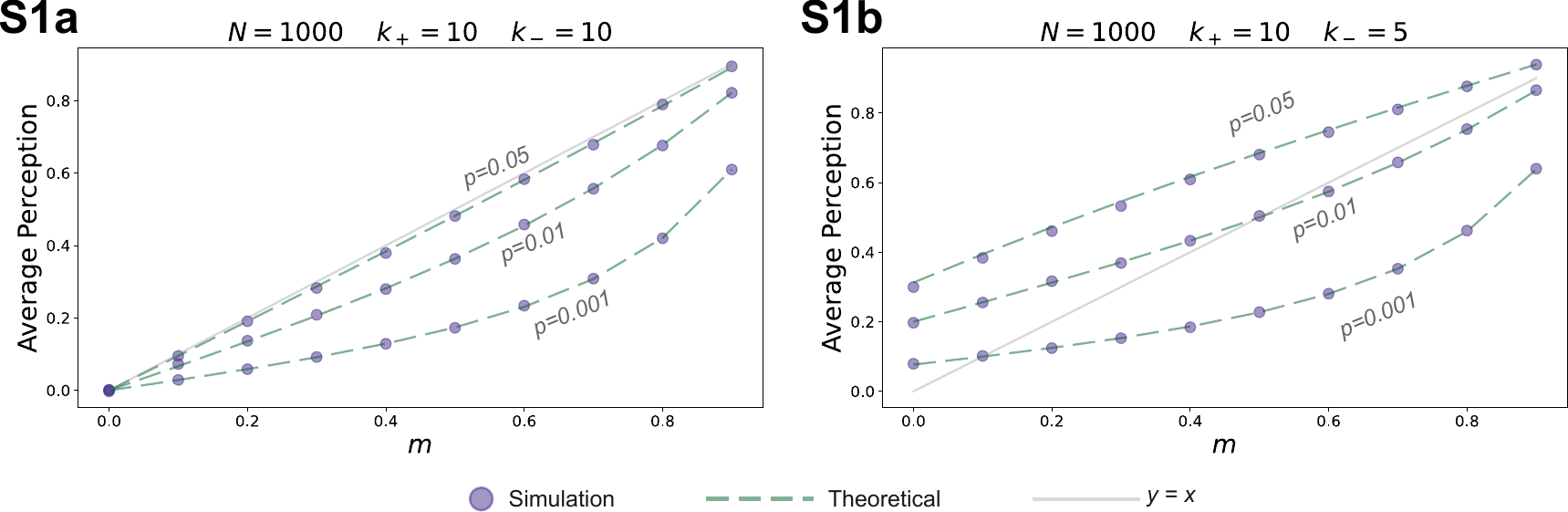}
    \caption{
    \textbf{Stationary state in simulations vs. theoretical solution.}
    The plots compare the average perception of the mean attribute $\mu$ reached in the simulations and compare them with the theoretical solution, as a function of $m$ and different values of $p$, both for $k_+=k_-$ (S1a) and $k_+ \neq k_-$ (S1b). The points for the simulations are averaged on $10$ runs each.
    }
    \label{fig:s3}
\end{figure}

Figure \ref{fig:s3} shows the perfect match between the stationary state of $\mu(d)$ in simulations and the theoretical solution $\mu_\infty$ of eq. (3) in the main text, obtained with community mean field. It also shows the difference between the average perception of the mean attribute and the mean attribute itself, highlighting how perception bias depends on the community size imbalance, connectivity homogeneity and polarization.

\section{Derivation of eq. (6) of the main text}

We start form eq. (3) of the main text for the stationary value of the perception according to community mean field calculations:
\begin{equation}
    \mu_{\infty} = \E \Big[\lim_{d\to\infty}\mu(d)\Big]=\frac{k_+N_+ - k_-N_-}{k_+N_+ + k_-N_- +4N_+N_-p}
\end{equation}
This expression can be conveniently rewritten in terms of the number of edges of the two communities:
\begin{equation}
    \mu_{\infty} = \frac{(E_++E_{c})-(E_-+E_{c})}{(E_++E_{c})+(E_-+E_{c})}
    \label{eq:m_th_edges}
\end{equation}
where $E_{\pm}$ are the edges within the $\pm1$ community and $E_{c}$ are the edges across the two communities. The stationary perception is then simply given by the number of messages sent by the $+1$ community minus the ones sent by the $-1$ community divided by the total number of messages. 

With this approximation, we can also write the expected values of the social circle of the two communities $\mu_\pm(1)$ in terms of $E_\pm$ and $E_c$, namely

\begin{equation}
\E[\mu_+(1)] = \frac{2E_+ - E_c}{2E_+ + E_c}, 
\qquad 
\E[\mu_-(1)] = -\frac{2E_- - E_c}{2E_- + E_c}.
\label{eq_mu_pm}
\end{equation}

For simplicity, we drop the expected value operator $\E[\quad]$.
Our goal is to express
\[
\frac{E_+ - E_-}{E_+ + E_- + E_c}
\]
in terms of \(\mu_+(1)\) and \(\mu_-(1)\).

---

\subsection*{Relations between \(\mu_+(1)\) and \(\mu_-(1)\).}

By direct algebraic manipulation,
\[
\mu_+(1) + \mu_-(1) 
= \frac{4E_c (E_+ - E_-)}{(2E_+ + E_c)(2E_- + E_c)},
\qquad
\mu_+(1) - \mu_-(1) 
= \frac{2(4E_+E_- - E_c^2)}{(2E_+ + E_c)(2E_- + E_c)}.
\]
Let the common denominator be
\[
D = (2E_+ + E_c)(2E_- + E_c).
\]

---

\subsection*{Express \(E_+ - E_-\) and \(E_+E_-\) in terms of \(\mu_+(1)\), \(\mu_-(1)\), and \(E_c\).}

From the first identity,
\[
E_+ - E_- = \frac{(\mu_+(1) + \mu_-(1))\,D}{4E_c}.
\]
From the second,
\[
4E_+E_- = \frac{(\mu_+(1) - \mu_-(1))\,D}{2} + E_c^2.
\]

---

\subsection*{Express \(D\) in terms of the total number of edges.}

Expanding \(D\),
\[
D = 4E_+E_- + 2E_c(E_+ + E_-) + E_c^2.
\]
Substitute the expression for \(4E_+E_-\):
\[
D = \frac{(\mu_+(1) - \mu_-(1))\,D}{2} + E_c^2 + 2E_c(E_+ + E_-) + E_c^2.
\]
Simplifying,
\[
D\left(1 - \frac{\mu_+(1) - \mu_-(1)}{2}\right)
= 2E_c(E_+ + E_- + E_c).
\]
Denoting the total number of edges by \(E = E_+ + E_- + E_c\), we obtain
\[
D = \frac{4E_c E}{\,2 - \mu_+(1) + \mu_-(1)\,}.
\]

---

\subsection*{Substitute back to eliminate \(D\).}

Using the earlier expression for \(E_+ - E_-\),
\[
\frac{E_+ - E_-}{E}
= \frac{(\mu_+(1) + \mu_-(1))D}{4E_c E}
= \frac{\mu_+(1) + \mu_-(1)}{\,2 - \mu_+(1) + \mu_-(1)\,}.
\]

---

\subsection*{Final identity}

\begin{equation}
\mu_\infty
= 
\frac{\mu_+(1) + \mu_-(1)}{\,2 - \mu_+(1) + \mu_-(1)\,}.
\end{equation}

\section{Fisher transformation}

Since the quantities under investigation (the perception estimates) are bounded in the interval $[-1, 1]$, their distribution becomes increasingly non-Gaussian as values approach the bounds. To perform statistical comparisons and uncertainty propagation in an approximately normal space, we applied the Fisher $z$-transformation \cite{Fisher1915}:

\[
z = \tanh^{-1}(r) = \frac{1}{2} \ln\!\left(\frac{1 + r}{1 - r}\right),
\]

where $r$ is the original bounded quantity. In this transformed space, the distribution of $z$ is approximately normal, particularly for $|r| < 0.9$, and the uncertainty propagates as

\[
\sigma_z = \frac{\sigma_r}{1 - r^2}.
\]

All statistical analyses were therefore performed in the Fisher $z$-space to ensure validity of the assumptions of approximate normality and additive uncertainties.

\section{Uncertainty of the estimators}
\label{si_unc}

In the main text, we use the theoretical stationary state and the social circle as predictors of the average perception of $m$ in the surveys. We recall the definitions of these two estimators

\begin{equation}
    \hat{\mu}_\infty= \frac{\overline{y}_+ + \overline{y}_-}{2- \overline{y}_+ + \overline{y}_-}
\qquad
    \hat{\mu}_1 = \overline{y}
\end{equation}

where, for each person interviewed in the survey, we have their own declared attribute $x_i$, their social circle $y_i$ and their perception of the mean attribute $z_i$. The overline denotes the sample mean, and the '$\pm$' subscript whether it is computed over one of the two communities.
\\
For the average perception, we estimate it as

\begin{equation}
    \hat\mu = n_+\overline{z}_+ + n_-\overline{z}_-
\end{equation}

whose uncertainty is simply given by the first-order propagation of the errors on the subsamples

\begin{equation}
    s^2(\hat{\mu}) = n_+^2 \frac{\sigma^2 (\overline{z}_+)}{\sqrt{S_+}} +n_-^2 \frac{\sigma^2 (\overline{z}_-)}{\sqrt{S_-}}
\end{equation}

where $\sigma^2 (\overline{z}_\pm)$ are the variances on the sub-samples and $S_\pm$ are the sizes of the sub-samples.

\subsection{Uncertainty on $\hat\mu_1$}
We start deriving the uncertainty on $\hat{\mu}_1$.\\
In our model $\mu_+(1)$ and $\mu_-(1)$ are correlated. This can also be inferred by equation \eqref{eq_mu_pm}, as the edges across the communities $E_c$ introduce a relation between the variables. Therefore, we assume $\overline{y}_+$ and $\overline{y}_-$ to be correlated, that the uncertainty must be computed over these values and then propagated into $\hat{\mu}_1$. As this estimator is the sample mean of $y$, we can rewrite it as

\begin{equation}
    \hat{\mu}_1 = n_+\overline{y}_+ + n_-\overline{y}_-
\end{equation}

where $n_\pm = \frac{N_\pm}{N}$. We assume to exactly know $n_\pm$, as $m$ is given in the survey, while the uncertainty to propagate comes from $\overline{y}_\pm$.
\\
The uncertainty on $\overline{y}_\pm$ is obtained through the standard deviation on the $\pm$ subsamples, divided by the square root of the size of the subsamples, thus

\begin{equation}
    s(\overline{y}_\pm) = \frac{\sigma(\overline{y}_\pm)}{\sqrt{N_{S_\pm}}}
\end{equation}

indicating with $N_{S_\pm}$ the subsample sizes.
We can easily use the formula for the propagated uncertainty on the social circle
\begin{equation}
    s^2(\hat{\mu}_1) = n_+ s^2(\overline{y}_+) + n_- s^2(\overline{y}_-) + 2n_+n_-\operatorname{Cov}(\overline{y}_+, \overline{y}_-)
    \label{eq:sc_sample}
\end{equation}

We now need to compute the measured covariance $\operatorname{Cov}(\overline{y}_+, \overline{y}_-)$. To do it, we compute the fluctuations deriving from $E_c$, affecting both $\mu_+$ and $\mu_-$ theorethically

\[
\operatorname{Cov}(\mu_+, \mu_-) \approx \frac{\partial\mu_+}{\partial{E_c}}\frac{\partial\mu_-}{\partial{E_c}}\sigma^2(E_c)
\]

\subsubsection*{Derivatives with respect to $E_c$}

Differentiate both quantities with respect to $E_c$:
\[
\frac{\partial \mu_+}{\partial E_c}
= -\frac{4E_+}{(2E_+ + E_c)^2},
\qquad
\frac{\partial \mu_-}{\partial E_c}
= \frac{4E_-}{(2E_- + E_c)^2}.
\]

\subsubsection*{Cross-term covariance approximation}

Using a first-order (delta-method) expansion, keeping only the shared dependence on $E_c$,
\[
\mathrm{Cov}(\mu_+, \mu_-)
\approx 
\frac{\partial \mu_+}{\partial E_c}
\frac{\partial \mu_-}{\partial E_c}
\,\sigma^2(E_c)
= 
-\frac{16\,E_+ E_-}{(2E_+ + E_c)^2 (2E_- + E_c)^2}\,\sigma^2(E_c).
\]

The negative sign arises because $\mu_+$ decreases with increasing $E_c$ 
while $\mu_-$ increases, so the two move in opposite directions.

\subsubsection*{Express in terms of $\mu_+$ and $\mu_-$}

We can eliminate $E_+$ and $E_-$ by rewriting them in terms of $\mu_+$, $\mu_-$, and $E_c$.  
From the definitions,
\[
\frac{E_+}{E_c} = \frac{1 + \mu_+}{2(1 - \mu_+)},
\qquad
\frac{E_-}{E_c} = \frac{1 - \mu_-}{2(1 + \mu_-)},
\]
and hence
\[
2E_+ + E_c = \frac{2E_c}{1 - \mu_+},
\qquad
2E_- + E_c = \frac{2E_c}{1 + \mu_-}.
\]

Substituting these into the covariance expression and simplifying gives:
\[
\operatorname{Cov}(\mu_+,\mu_-)
\approx 
-\,\frac{(1 - \mu_+^2)(1 - \mu_-^2)}{4\,E_c^{\,2}}\,\sigma^2(E_c).
\]

\subsubsection*{Equivalent forms using variances of $\mu_+$ or $\mu_-$}

Using the delta-method approximations
\[
\sigma^2(\mu_+) \approx 
\left(\frac{\partial \mu_+}{\partial E_c}\right)^2 \sigma^2(E_c),
\qquad
\sigma^2(\mu_-) \approx 
\left(\frac{\partial \mu_-}{\partial E_c}\right)^2 \sigma^2(E_c),
\]
we can eliminate $\sigma^2(E_c)$ to obtain purely observable forms:
\[
\operatorname{Cov}(\mu_+,\mu_-)
\approx
-\,\frac{1 - \mu_-^2}{1 - \mu_+^2}\,\sigma^2(\mu_+)
=
-\,\frac{1 - \mu_+^2}{1 - \mu_-^2}\,\sigma^2(\mu_-).
\]

This formula can be conveniently written as the geometric mean of the last two terms equaling $\sigma^2(\mu_+,\mu_-)$

\begin{equation}
    \operatorname{Cov}(\mu_+,\mu_-) \approx \sigma(\mu_+)\sigma(\mu_-)
    \label{eq:cov}
\end{equation}

The covariance between the social circles of the two communities is thus positive, because an increase of one of the two towards $+1$ implies more in the $+1$ community or a higher connectivity of those nodes, which also reflects on the other community's social circle.
\\
\\
Plugging this value into equation \eqref{eq:sc_sample}, and treating the sampled values as the theoretical ones, we get 

\begin{equation}
    s(\hat{\mu}_1) = n_+ s(\overline{y}_+) + n_- s(\overline{y}_-)
    \label{eq:sc_sample_err}
\end{equation}

as the final term completes the square.

\subsection{Uncertainty on $\hat\mu_\infty$}
For compactness set
\[
N \equiv \mu_+ + \mu_-, \qquad D \equiv 2 - \mu_+ + \mu_-,
\qquad \text{so}\qquad \mu_{\infty}=\frac{N}{D}.
\]

after we dropped $(1)$ from $\mu_\pm(1)$.

\subsubsection*{Partial derivatives (quotient rule)}

Using the quotient rule
\(\displaystyle \frac{\partial (N/D)}{\partial x}=\frac{(\partial N/\partial x)\,D - N\,(\partial D/\partial x)}{D^2}\),

for \(\mu_+\):
\[
\frac{\partial \mu_{\infty}}{\partial \mu_+}
= \frac{(1)\,D - N(-1)}{D^2}
= \frac{D+N}{D^2}.
\]
Compute \(D+N\):
\[
D+N=(2-\mu_+ + \mu_-)+(\mu_+ + \mu_-)=2+2\mu_-=2(1+\mu_-),
\]
thus
\[
\frac{\partial \mu_{\infty}}{\partial \mu_+}
= \frac{2(1+\mu_-)}{(2-\mu_+ + \mu_-)^2}\; .
\]

For \(\mu_-\):
\[
\frac{\partial \mu_{\infty}}{\partial \mu_-}
= \frac{(1)\,D - N(1)}{D^2}
= \frac{D-N}{D^2}.
\]
Compute \(D-N\):
\[
D-N=(2-\mu_+ + \mu_-)-(\mu_+ + \mu_-)=2-2\mu_+ = 2(1-\mu_+),
\]
so
\[
\frac{\partial \mu_{\infty}}{\partial \mu_-}
= \frac{2(1-\mu_+)}{(2-\mu_+ + \mu_-)^2}\;
\]

\subsubsection*{First-order propagated variance}

Let \(s^2(\mu_+)\) and \(s^2(\mu_-)\) denote the sampled variances of \(\mu_+\) and \(\mu_-\),
and \(\operatorname{Cov}(\mu_+,\mu_-)\) their covariance. The linear propagation formula yields
\[
\begin{aligned}
s^2(\mu_{\infty})
&= \left(\frac{\partial \mu_{\infty}}{\partial \mu_+}\right)^{\!2} s^2(\mu_+)
+ \left(\frac{\partial \mu_{\infty}}{\partial \mu_-}\right)^{\!2} s^2(\mu_-) \\
&\qquad + 2\left(\frac{\partial \mu_{\infty}}{\partial \mu_+}\right)
  \left(\frac{\partial \mu_{\infty}}{\partial \mu_-}\right)
  \operatorname{Cov}(\mu_+,\mu_-).
\end{aligned}
\]

Substituting the derivatives:
\[
\begin{aligned}
s^2(\mu_{\infty})
&= \frac{4(1+\mu_-)^2}{(2-\mu_+ + \mu_-)^4}\,s^2(\mu_+)
+ \frac{4(1-\mu_+)^2}{(2-\mu_+ + \mu_-)^4}\,s^2(\mu_-) \\
&\qquad
+ 2\cdot\frac{2(1+\mu_-)}{(2-\mu_+ + \mu_-)^2}\cdot
\frac{2(1-\mu_+)}{(2-\mu_+ + \mu_-)^2}\,
\operatorname{Cov}(\mu_+,\mu_-) \\
&= \frac{4}{(2-\mu_+ + \mu_-)^4}
\Big[ (1+\mu_-)^2\,s^2(\mu_+) + (1-\mu_+)^2\,s^2(\mu_-) \\
&\qquad\qquad\qquad\qquad
+ 2(1+\mu_-)(1-\mu_+)\,\operatorname{Cov}(\mu_+,\mu_-)\Big].
\end{aligned}
\]

Therefore the propagated sampled-variance is
\[
s^2(\mu_{\infty})
= \frac{4}{(2-\mu_+ + \mu_-)^4}
\Big[ (1+\mu_-)^2\,s^2(\mu_+) + (1-\mu_+)^2\,s^2(\mu_-)
+ 2(1+\mu_-)(1-\mu_+)\,\operatorname{Cov}(\mu_+,\mu_-)\Big]. 
\]

By replacing the value for the covariance expressed in \eqref{eq:cov}, and using $\overline{y}_\pm$ to estimate $\mu_\pm$, we get the final formula

\begin{equation}
    s(\hat\mu_{\infty})
= \frac{2}{(2-\overline{y}_+ + \overline{y}_-)^2}
\Big[ (1+\overline{y}_-)\,s(\overline{y}_+) + (1-\overline{y}_+)\,s(\overline{y}_-)
\Big].
\end{equation}

\section{Model comparison via $\chi^2$ statistics}

To quantify how well each estimator reproduced the empirical (survey-based) values, we computed a reduced chi-squared ($\chi^2_\nu$) statistic for each country and estimator. This metric measures the average deviation between predicted and observed values in units of the combined uncertainty, thus providing a scale-independent measure of model performance. Specifically, for $N$ issues within a country, the statistic was computed as

\[
\chi^2 = \sum_{i=1}^{N} \frac{(E_i - T_i)^2}{\sigma_{E_i}^2 + \sigma_{T_i}^2},
\]

where $E_i$ and $\sigma_{E_i}$ denote the estimated value and its propagated uncertainty, while $T_i$ and $\sigma_{T_i}$ represent the corresponding empirical (true) value and its measurement uncertainty.  
The reduced chi-squared is then given by

\[
\chi^2_\nu = \frac{\chi^2}{\nu},
\]

with $\nu = N$ degrees of freedom, as the estimators are based on the own attribute $x$ and the social circle $y$ declared by the users, and then we fit $\overline{z}$.

A value of $\chi^2_\nu \approx 1$ indicates that the deviations between predicted and observed values are consistent with their uncertainties, while $\chi^2_\nu \gg 1$ suggests that the estimator systematically underestimates the uncertainty or poorly fits the data. Conversely, $\chi^2_\nu \ll 1$ implies overestimated uncertainties or overfitting.


\section{Label Propagation}

This algorithm is used to predict missing attributes in a network, with a path-dependent message-passing that uses the known labels \cite{Zhu2002}.
Instead, in our graphs we have all the attributes, and interpret the predicted label for the node as its perception about $m$.
The mechanism is similar to DeGroot, but it also considers how many paths lead from one node to the other. Therefore, we first compute the matrix $T_{ij}$, which gives the probability of going from node $i$ to node $j$ with a path of length $1$, namely
\begin{equation}
    \mathbf{T} = \mathbf{D^{-1}}\mathbf{A}
\end{equation}
where $\mathbf{D}$ is the degree matrix.
Then, we can easily get the probability matrix to go from node $i$ to node $j$ with a random path of lenght $d$, by simply computing $\mathbf{T}^d$.
Now we just need to consider the attributes vector $s_i$, and we can get the propagated attributes vector (nodes perceptions) at distance $d$, by computing

\begin{equation}
    \boldsymbol{\mu}(d) = \mathbf{T}^d \mathbf{s}
\end{equation}

\subsubsection*{Comparison with DeGroot model}
Both methods aggregate the nodes' neighbors' perceptions and iterate the process to distance $d$.
In these models, the own attribute of the node is propagated at $d=1$, and partially "bounces back" to the node at $d=2$, because its neighbors' perceptions depend on the node we are considering. However, our tests confirm this effect to be negligible, and, anyway, it may not be considered wrong, as it can be realistic to be influenced by our peers, who, in turn, are influenced by ourselves.
Label propagation is more complex than DeGroot, as it also considers the number of paths leading from node $i$ to node $j$.
However, both methods lead to the same stationary solution. The key difference is that Label propagation gets to it faster than DeGroot.
Since this is just a minor difference between the two models, the plots in the main text are obtained through DeGroot.

\subsection*{Simulations with Label Propagation on SBM and Karate club}

We repeat the simulations reported in the main text with Label Propagation, finding no significant differences with DeGroot model. The evolution is faster, but the stationary state is the same for both communities and overall.

\begin{figure}[H]
    \centering
    \includegraphics[width=0.98\linewidth]{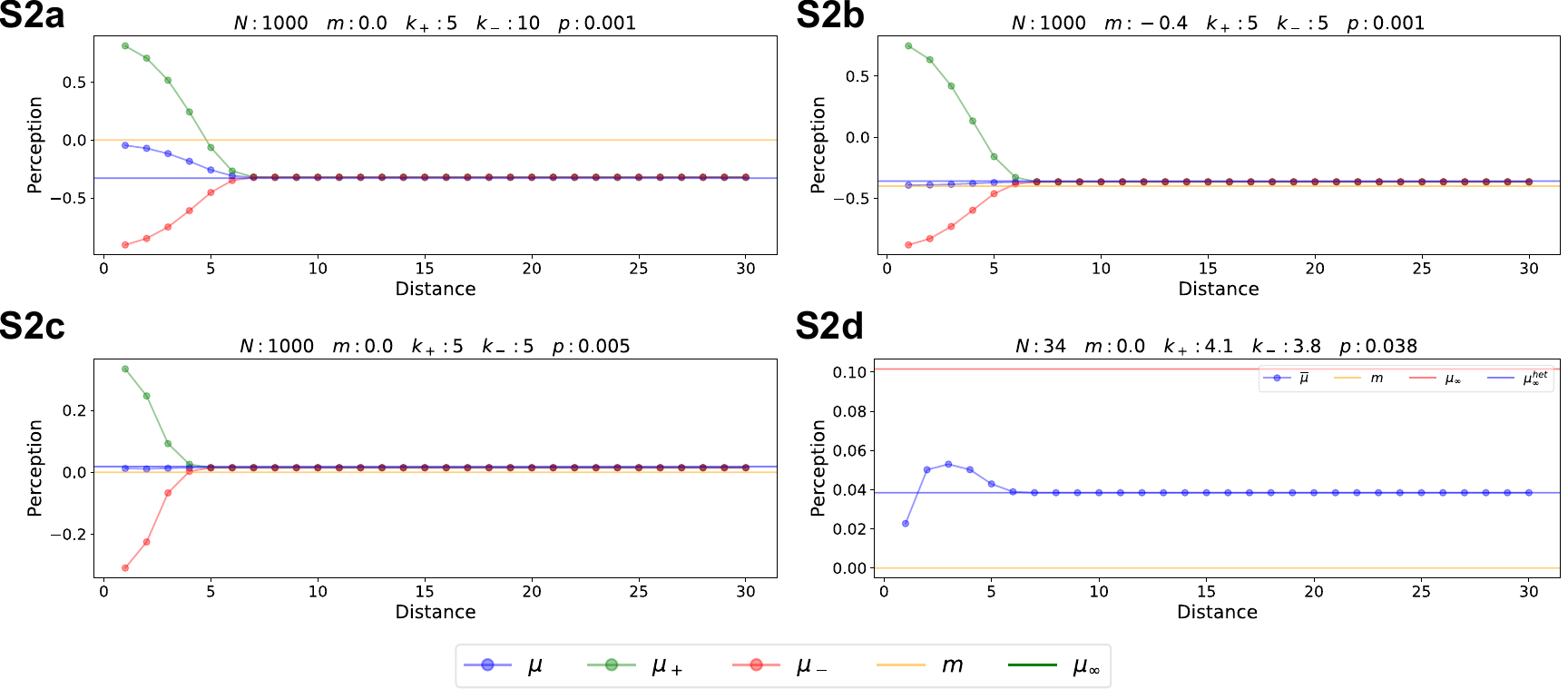}
    \caption{\textbf{Simulations with Label Propagation on SBM and Karate Club network}\\
    2a) SBM with same community size but different connectivity of communities;\\
    2b) SBM with same connectivity but different community size;\\
    2c) SBM with same community size and connectivity with high mixing (low polarization);\\
    2d) Karate club graph with affiliation attribute. \\
    The plots show no significant difference with the simulations with DeGroot model, except for the evolution being faster with label propagation.}
    \label{fig:s2}
\end{figure}

\section{Homophilic Barabási–Albert (HBA) graph}

Lee et al. \cite{Lee2019} introduced a modified version of the Barabási–Albert model \cite{Barabasi1999}. 
They modified the preferential-attachment process by adding a homophily parameter $q$, which adjusts the probability that a new node connects to others depending on whether they share the same attribute $s$. 
In this way the graph naturally separates into communities of nodes with the same attribute, while keeping a scale-free degree distribution. For full details on the model we remand to the original paper \cite{Lee2019}.

However, this model is less suitable for our study than SBM, because we cannot control the average degree of the two communities—with important consequences. If we have two communities with different size, nodes of the larger community are more likely to enter the graph before the others; as a consequence of the preferential attachment process they end up with higher degrees. Hence the average degree of the larger community is always higher and cannot be tuned: the generated graphs always fall in the case of Fig. 2a of the main text, in which perception bias is caused by different connectivity of the communities. With HBM we cannot explore different scenarios, at stake with SBM.

\begin{figure}[H]
    \centering
    \includegraphics[width=\linewidth]{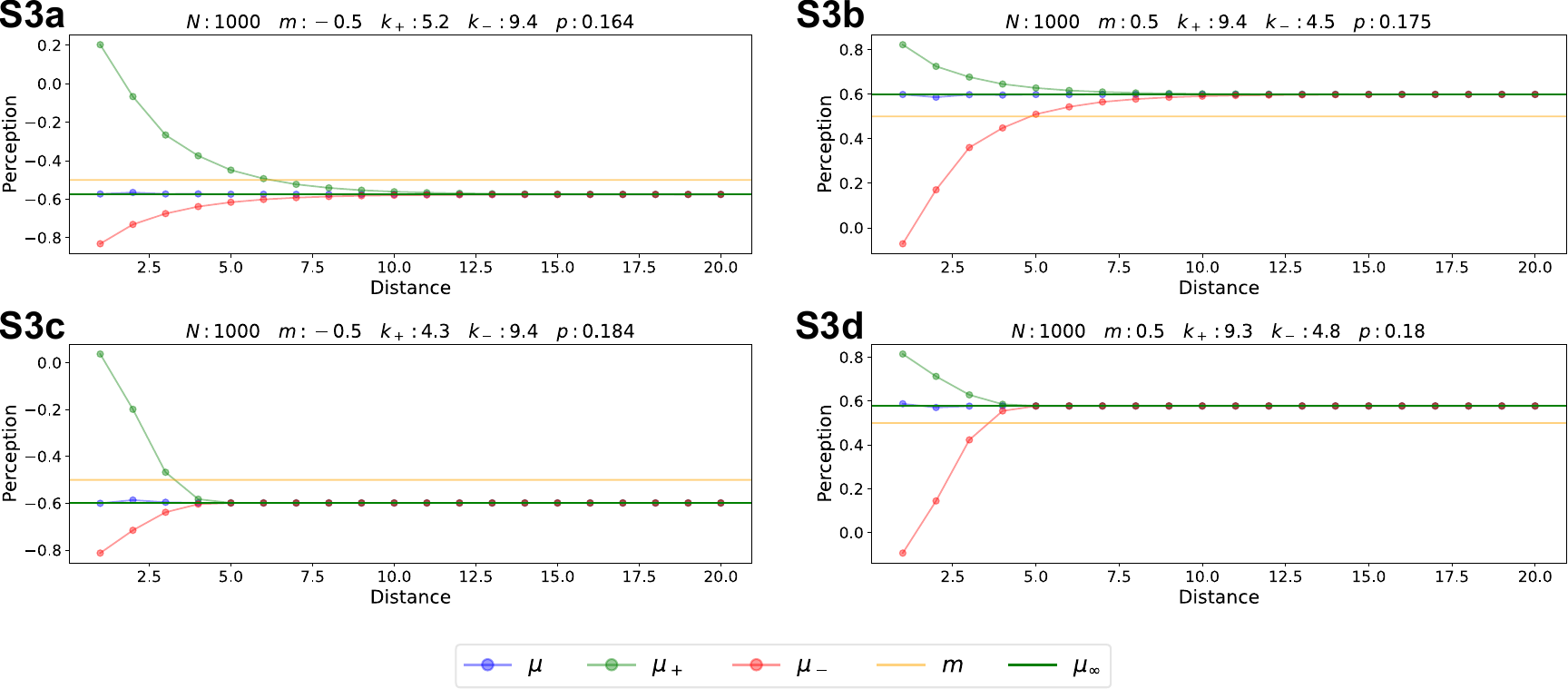}
    \caption{\textbf{Simulations with De Groot model on HBA}\\
    Each plot shows a simulation of the perceived mean attribute on a different network with size $N=1000$, for negative $m$ (2a and 2c) with different average community degree, and similarly for positive $m$ (2b and 2d). The evolution of $\mu$ as a function of $d$ is similar to SBM, and out theoretical solution is still a good predictor of the stationary state.
    }
    \label{fig:s1}
\end{figure}

Simulations results for HBA, reported in Figure \ref{fig:s1}, confirm that our findings hold also in this setup.

\end{document}